# New method of $^{85}$Kr reduction in a noble gas based low-background detector

D.Yu. Akimov[a,b,*], A.I. Bolozdynya[b], A.A. Burenkov[a,b], C. Hall[c], A.G. Kovalenko[a,b], V.V. Kuzminov[d], G.E. Simakov[a]

[a] *SSC RF Institute for Theoretical and Experimental Physics of National Research Centre "Kurchatov Institute", 25 Bolshaya Cheremushkinskaya, Moscow, 117218, Russian Federation*

[b] *National Research Nuclear University MEPhI (Moscow Engineering Physics Institute), 31 Kashirskoe highway, Moscow, 115409, Russian Federation*

[c] *University of Maryland, College Park, MD 20742, USA*

[d] *Institute for Nuclear Research RAS, 7a prospekt 60-letiya Oktyabrya, Moscow 117312, Russian Federation*

   *E-mail:* akimov_d@itep.ru

ABSTRACT: Krypton-85 is an anthropogenic beta-decaying isotope which produces low energy backgrounds in dark matter and neutrino experiments, especially those based upon liquid xenon. Several technologies have been developed to reduce the Kr concentration in such experiments. We propose to augment those separation technologies by first adding to the xenon an $^{85}$Kr-free sample of krypton in an amount much larger than the natural krypton that is already present. After the purification system reduces the total Kr concentration to the same level, the final $^{85}$Kr concentration will have been reduced even further by the dilution factor. A test cell for measurement of the activity of various Kr samples has been assembled, and the activity of 25-year-old Krypton has been measured. The measured activity agrees well with the expected activity accounting for the $^{85}$Kr abundance of the earth's atmosphere in 1990 and the half-life of the isotope. Additional tests with a Kr sample produced in the year 1944 (before the atomic era) have been done in order to demonstrate the sensitivity of the test cell.

KEYWORDS: Noble liquid detectors (scintillation, ionization, double-phase), Dark Matter detectors (WIMPs, axions, etc.)

---

[*] Corresponding author.

# Contents

## 1. Introduction

Presence of the beta-decaying anthropogenic $^{85}$Kr isotope in xenon is one of the most serious problems in the xenon based dark matter WIMP search experiments. Although the abundance of this isotope in natural Krypton (natKr) is only ~$2 \cdot 10^{-11}$ and the concentration of krypton in xenon is typically rather small ranging from $10^{-9}$ to $10^{-6}$ mol/mol (natKr/Xe) depending on the manufacturer, $^{85}$Kr is, nevertheless, highly problematic for dark matter experiments because of the high decay rate ($T_{1/2}$=10.76 years). The $^{85}$Kr isotope is produced in nuclear power plants and is released into the atmosphere. Xenon gas is extracted from the atmosphere together with other noble gases including Krypton. Therefore, the presence of $^{85}$Kr in modern xenon is inevitable. The experimental groups searching for dark matter have developed methods to further purify $^{nat}$Kr from Xe: gas chromatography and distillation in a rectification column [1], [2]. These methods allow one to achieve the level of $^{nat}$Kr/Xe of several ppt ($10^{-12}$ mol/mol). Xenon with this purity level is used in the current generation of the dark matter experiments [3], [4], [5], [6]. For the future multi-ton dark matter detectors (LZ [7], for example), a much lower level of Kr concentration in Xe will be required. On the other hand, the concentration of ~$10^{-12}$ mol/mol is currently the lowest practically detectible value by mass-spectrometric method developed for a working detector (see [8], [9] for details of the method for the EXO and LUX experiments). As shown in [9], the Kr signal can be clearly identified at concentrations as low as $0.5 \cdot 10^{-12}$ mol/mol (natKr/Xe). We propose here a method to further reduce the $^{85}$Kr concentration of Xe by several orders of magnitude while using existing separation techniques.

## 2. Method

We may consider for simplicity the Kr removal system to be a device which acts upon the input xenon gas stream and decreases the Kr concentration to a fixed level. We refer to this type of purification system as Type-I. Then, the principle of our method is as follows. A sample of Kr depleted from the $^{85}$Kr isotope is added and diluted to the Xe. Then the Xe sample is purified of Kr with a Type-I system, returning it to the initial level or lower by one of existing methods. The natural Kr has been diluted with the depleted one and is removed together with it since both of them have practically the same thermodynamic properties. Let us introduce for the final mixture of the depleted and natural Kr an equivalent concentration $C_{Kr/Xe}^{eq}$ which would be equal to the concentration of natural Kr in Xe containing the same amount of $^{85}$Kr. This concentration is expressed by the formula:

$$C_{Kr/Xe}^{eq} = C_{Kr/Xe} \cdot \frac{q \cdot C_{depKr/Xe} + C_{Kr/Xe}^{ini}}{C_{depKr/Xe} + C_{Kr/Xe}^{ini}}, \qquad (2.1)$$

where $C_{Kr/Xe}$ is final Kr concentration in Xe achievable by standard physical methods, $C_{Kr/Xe}^{ini}$ is initial concentration of Kr in Xe before adding the depleted Kr, $C_{depKr/Xe} = M_{depKr}/M_{Xe}$ ($M$ is a molar mass) is the concentration of the depleted Kr in the Xe after mixing, and $q$ is depletion factor (<1). For ultimate case, if the depleted Kr sample is absolutely pure from $^{85}$Kr ($q$=0), then $C_{Kr/Xe}^{eq}$ will be defined as:

$$C_{Kr/Xe}^{eq} = C_{Kr/Xe} \cdot \frac{C_{Kr/Xe}^{ini}}{C_{depKr/Xe} + C_{Kr/Xe}^{ini}}. \qquad (2.2)$$

In the case when $C_{Kr/Xe}^{ini}$ is rather small with respect to $C_{depKr/Xe}$ the reduction factor is simply defined by the ratio $C_{Kr/Xe}^{ini}/C_{depKr/Xe}$. For example, if one ton of Xe has already undergone purification from Kr to the ppt level, i.e., it contains only ~1 μg of Kr, then applying this method with only 1 g of $^{85}$Kr-free Kr will result in $C_{Kr/Xe}^{eq}$ ~10$^{-18}$ mol/mol. This corresponds to only one $^{85}$Kr atom in approximately 10 tons of Xenon!

Krypton depleted from $^{85}$Kr may be obtained by centrifugation. The use of Kr enriched by the lightest stable $^{78}$Kr isotope is preferred because it must have the lowest abundance of $^{85}$Kr. The isotope abundance of Krypton enriched by $^{78}$Kr produced in Russia is shown in table 1 together with the isotope abundance of natural Krypton. One can see that the abundance of $^{84}$Kr, the most abundant isotope in the natural Kr, is only ~0.0002. Thus, the depletion factor ($q$) is 0.0002/0.57≈3.5·10$^{-4}$. For the $^{85}$Kr isotope the $q$ is expected to be even lower. Another possibility is to use a very old gas, produced before the atomic era (see below).

We may also consider an alternate model for the Kr removal system, which is closer to reality and which we call type-II. In this type of system, the Kr is not reduced to a fixed level, but instead on each pass is reduced by some known factor $R$; $N$ passes through the system then reduce the Kr concentration by a factor $R^N$. In this case and in the case of very thoroughly prepared gas system (i.e. without any internal sources of $^{nat}$Kr caused by leaks to the air, outgassing from the parts that were exposed to air etc.), the addition of the $^{85}$Kr-free krypton may allow the operator to observe the elevated $^{78}$Kr or $^{84}$Kr level and thereby to monitor the performance of the purification system and to finally asses the $^{85}$Kr level that corresponds to the $R^N$ reduction factor when the level of $^{78}$Kr or $^{84}$Kr have reached a detectible limit.

Table 1. Isotope abundance for natural and $^{78}$Kr-enriched Krypton.

| Isotope | 78 | 80 | 82 | 83 | 84 | 86 |
|---|---|---|---|---|---|---|
| Natural Kr [10] | 0.00356 | 0.0227 | 0.116 | 0.115 | 0.57 | 0.173 |
| Enriched $^{78}$Kr [11] | 0.9408 | 0.0588 | 0.0001 | 0.0001 | 0.0002 | - |

## 3. Measurement of Kr radioactivity

For estimation of $C_{Kr/Xe}^{eq}$ one must know the $q$ value. This can be done by measurement of the activity of the depleted Kr sample. We selected for this purpose a scintillation spectrometric method in a liquid phase. This method allows one to measure a large amount of Kr in a compact cell. We constructed a simple test chamber filled with condensed Kr in order to understand the accuracy of measurements which can be performed with it. The sketch of this chamber is shown in figure 1. The chamber is assembled from standard CF vacuum pieces. A liquid $N_2$ bath is used for cooling the chamber. The test cell is made of 2¾" CF nipple (35 mm inner diameter) with a blank flange on the bottom. The cell is viewed with a 1" $MgF_2$ window multi-alkali FEU-181 PMT (produced by MELZ, Moscow) hanging inside the tube at a height of 33 mm from the bottom. The PMT is sensitive to the 152-nm VUV luminescence of Kr. Scintillation signals detected by the PMT are shaped (with τ~1-2 μs) and multiplied by an ORTEC 570 amplifier and then analyzed by an ORTEC 927 MCA. Measurements were performed with two configurations of the test cell, see figure 2. In configuration #1 (in figure 2 on left), the whole space inside the 2¾" CF nipple was used, and the liquid Kr filled the cell to a level of just above the PMT photocathode. The amount of Kr condensed into the cell in configuration #1 is 77±5 g. In configuration #2 (in figure 2 on right), a stainless steel insert with an inner diameter of ~20 mm and "black walls" made by machining an M20 thread was added to the cell in order to improve its spectrometric properties. An $^{241}$Am alpha-source was installed in the middle of the cell at a distance of 5-mm from the bottom. The amount of condensed Kr in configuration #2 was 5.6±0.2 g, corresponding to a liquid level of ~8.5 mm above the cell bottom. Two samples of Kr with different $^{85}$Kr activity were tested in the cell. The first was produced in the year 1990, and the second one, in the year 1944. The latter one was actually a mixture with 7% Xe; however, we consider that the Xe did not make any significant influence to our measurements, because it was probably frozen on the cell walls. The reason to use the 1944 $^{85}$Kr-free sample is to measure the own background count rate of the cell.

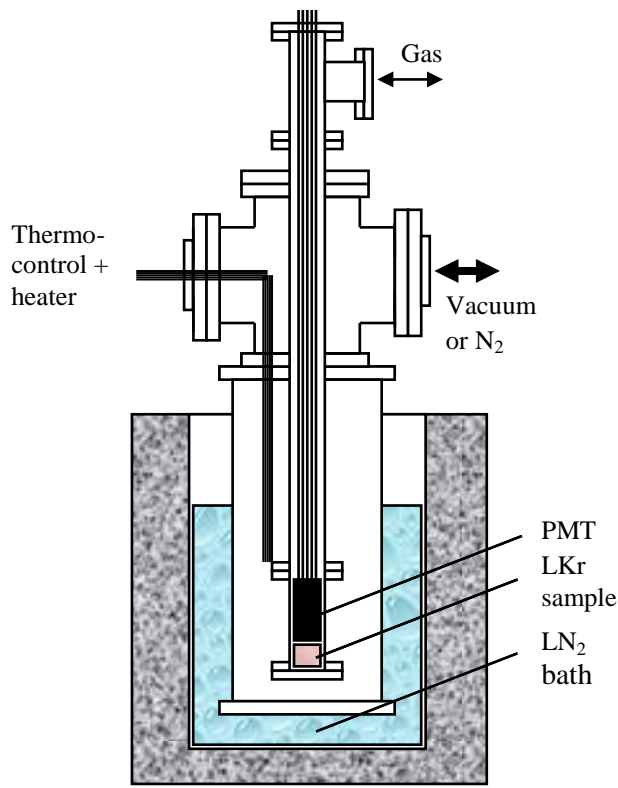

**Figure 1.** Schematic diagram of the test chamber.

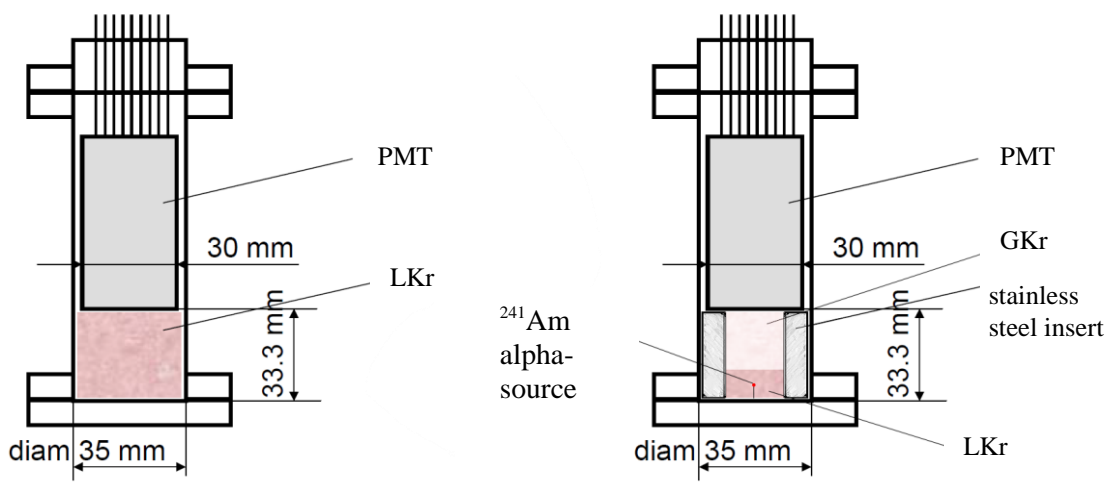

**Figure 2.** Schematic view of two test cell configurations.

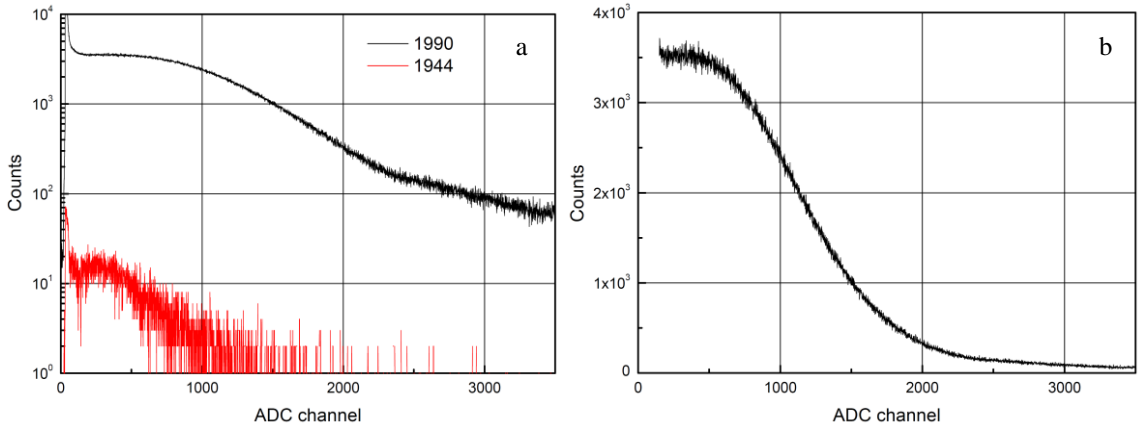

**Figure 3.** Pulse height spectra obtained with the test cell in configuration #1; a – spectra measured with 1990-y and 1944-y Kr samples; b – spectrum obtained by subtraction of the 1944-y Kr spectrum from 1990-y Kr spectrum.

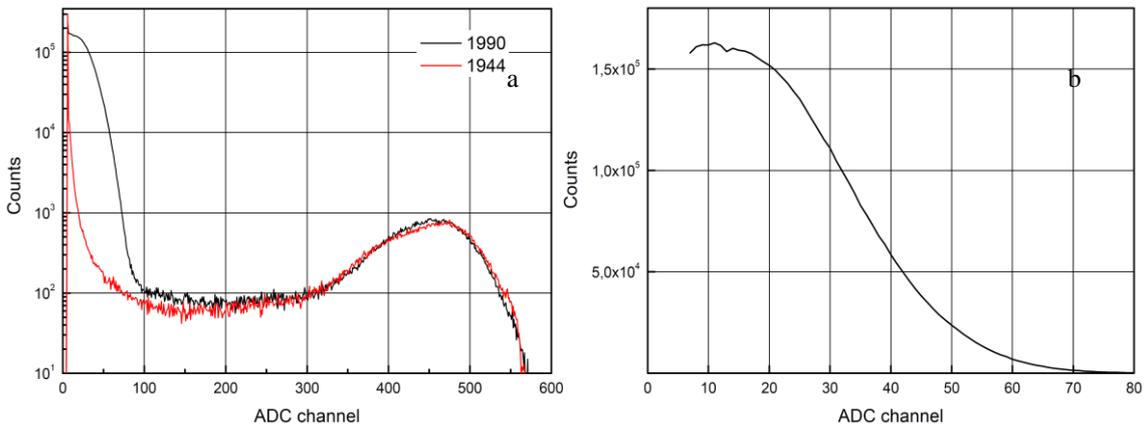

**Figure 4.** Pulse height spectra obtained with the test cell in configuration #2; a – spectra measured with 1990-y and 1944-y Kr samples, b – spectrum obtained by subtraction of 1944-y Kr spectrum from 1990-y Kr spectrum.

Pulse height spectra measured in configuration #1 for the 1990-y and 1944-y samples of Kr are shown in figure 3a. The duration of the measurements was 1000 live seconds. The spectrum shown in figure 3b is obtained by subtraction of the 1944-y Kr spectrum from the 1990-y Kr spectrum, and thus, is considered to be related only to the activity of $^{85}$Kr. Similar pulse height spectra measured in configuration #2 (during 10,000 live seconds) are shown in figure 4. During measurements with the 1944-y sample of Kr (in configuration #2), the PMT

– 4 –

HV was slightly reduced to have the alpha-peak position at the same place as in the measurements with the 1990-y sample.

The spectra shown in figure 3b and figure 4b are convolutions of a true beta-decay spectrum and the cell response functions. Nevertheless, they give us the possibility to estimate the total count rates of the samples, assuming in first approximation a flat behavior of the spectra in the low-energy region (<150 ch in figure 3b and <7 ch in figure 4b). Such an approximation is valid because according to [12], the shape of the $^{85}$Kr beta-spectrum in the low-energy region is practically flat. The obtained values are 56±5 and 51±3 Bq/g for measurements in configurations #1 and #2, respectively. Extrapolated back to the year 1990 (with $T_{½}$=10.76 y), the activity of Kr at that time was 280±25 and 255±15 Bq/g, respectively. Taking into account the content of Kr in the air of 1.14 ppm, one can obtain the radioactivity of the air due to $^{85}$Kr decays of 1.20±0.10 and 1.10±0.06 Bq/m$^3$ for the measurements in configurations #1 and #2, respectively. These values are very close to the average air radioactivity due to $^{85}$Kr in Europe at that time: see figure 5 from Ref. [13] showing the data from the Mount Schauinsland monitor in Germany; the obtained points are superimposed on this plot. This demonstrates the correctness of the described method of Kr radioactivity measurement. The ultimate sensitivity of the test cell may be estimated from the spectra presented in figure 3a and figure 4a. From figure 3a, one can see that the maximum count rate ratio of two spectra reaches ~10$^3$ in the region around channel 1000 of the ADC. For figure 4a this ratio is ~10$^2$ in the region of ADC channels from ~30 to ~60. However, there are substantial substrates in this region from the alpha-source. Without these substrates, the ratio can be estimated at the same level of ~$10^3$. Thus, we may claim that in both configurations the radioactivity of sample caused by $^{85}$Kr can be reliably measured at the same level of magnitude as the cell background (i.e. ~$10^{-3}$ of the current radioactivity of the 1990-y Kr sample) provided the cell background is measured beforehand with $^{85}$Kr-free krypton. The present day average activity of the air due to $^{85}$Kr is stable since the year 2000 and is equal to ~1.45 Bq/m$^3$ [13]. Newly produced Kr from air would have an activity of ~340 Bq/g that is by a factor of 6.07±0.6 greater than the activity of our 1990-y sample. Thus, we may say that the radioactivity of present day Kr depleted from $^{85}$Kr with a factor q=1.6·$10^{-4}$ (approximately the same as that expected with centrifuging) is reliably measurable.

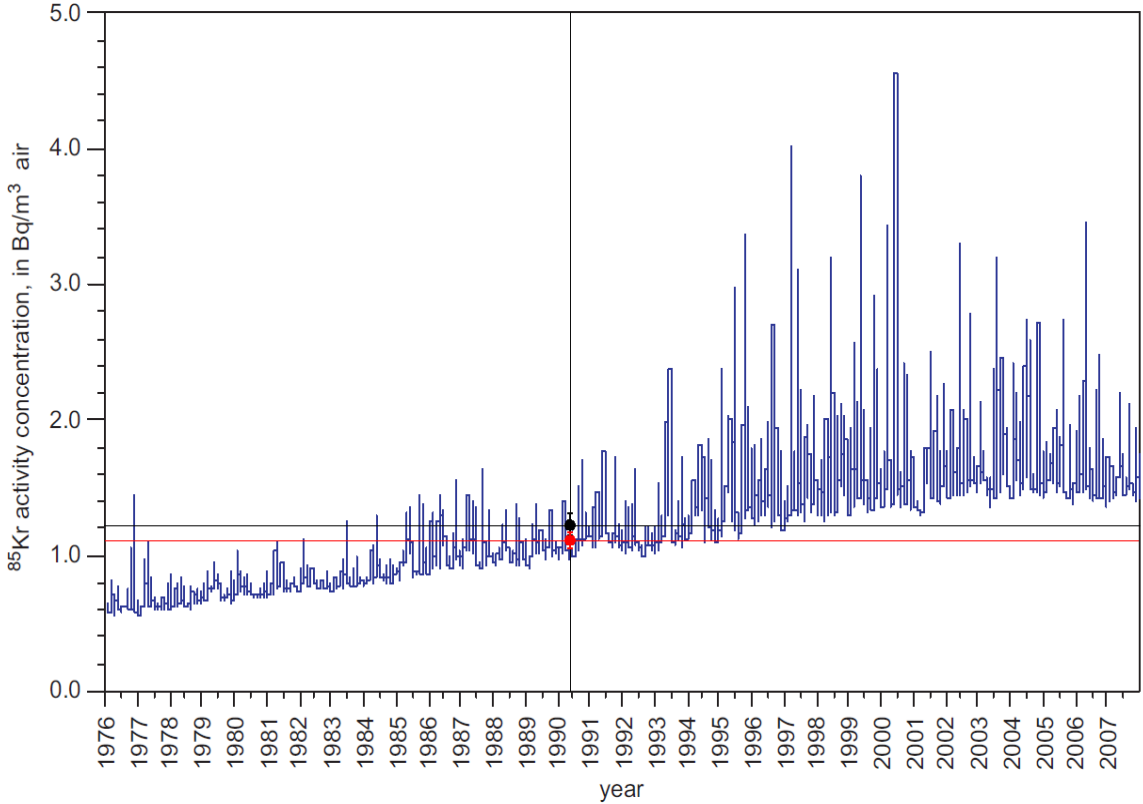

**Figure 5.** Comparison of the obtained radioactivity of the air due to $^{85}$Kr with the data from Mount Schauinsland monitor [13]; black and red circles are the data obtained with configuration #1 and #2, correspondingly.

## 4. Conclusion

We propose a new method of $^{85}$Kr reduction in a noble gas low-background detector. The method is based upon adding to the detector medium an $^{85}$Kr-free Krypton sample in an amount much larger than the initial natural Krypton content and subsequent reduction of the Kr concentration down to the initial or lower level with the use of existing methods of Xe-Kr separation. This method allows one to reduce the content of $^{85}$Kr in a detector medium by several orders of magnitude with respect to that achievable with the known methods. It has been shown with the assembled test cell that the radioactivity of $^{85}$Kr-free Krypton sample can be measured correctly by scintillation spectrometry in a liquid (condensed) phase. The measured activity recalculated (taking into account $^{85}$Kr life time) to the air activity in the year 1990 is very close to the available data on the air activity at that time. A test with a Kr sample produced in the year 1944 (before the atomic era) has been done in order to define the test cell sensitivity. It has been demonstrated that the level of $^{85}$Kr of ~1.6·$10^{-4}$ from the present day content of $^{85}$Kr in natural Kr is reliably detectable.

## Acknowledgments

We are grateful to I.R. Barabanov for the very valuable discussion on the proposed method.